# Scrapyard AI

[1] Marc Böhlen
marcbohlen@protonmail.com
[2] Sai Krishna
saikrishna12377@gmail.com

**Abstract.** This paper considers AI model churn as an opportunity for frugal investigation of large AI models. It describes how the incessant push for ever more powerful AI systems leaves in its wake a collection of obsolete yet powerful AI models, discarded in a veritable scrapyard of AI production. This scrapyard offers a potent opportunity for resource-constrained experimentation into AI systems. As in the physical scrapyard, nothing ever truly disappears in the AI scrapyard, it is just waiting to be reconfigured into something else.

Project Nudge-x is an example of what can emerge from the AI scrapyard. Nudge-x seeks to manipulate legacy AI models to describe how mining sites across the planet are impacting landscapes and lives. By sharing this collection of brutal landscape interventions with people and AI systems alike, Nudge-x creates a venue for the appreciation of a history sadly shared between AI and people.

## 1 Introduction

Planetary computing is expanding in intensity and scope. While the fury toward industrial scale Artificial Intelligence (AI) continues unabated, a new reality has materialized in the shadows of this first condition. As each new model replaces a previous AI model at increasingly faster rates, a literal scrapyard of AI models is amassed, not unlike junkyards of earlier industrial products, such as automobiles and household appliances. Only now, the time to reach decommission and obsolescence is much shorter. Not decades, not years, but months and sometimes even shorter times.

This paper casts Scrapyard AI as a new technical-cultural condition and then describes a media project, Nudge-x, which responds specifically to the condition of AI obsolescence by crafting new operations from a collection of older AI models. Nudge-x is a planetary computing intervention that nudges scrapyard AI systems to gain an understanding of the impact of industrial level mining operations on landscapes across Earth.

## 2 AI Obsolescence

In the current battle for AI supremacy, large technology companies are releasing increasingly powerful and increasingly costly AI models. OpenAI released GPT-5 [OpenAI 2025], Anthropic released Claude 3.5 and 4 [Anthropic 2025], Google released fourteen different Gemini models since April 2025 [DeepMind-Gemini 2025], [DeepMind-Modelcards 2026]. Meta released Llama 4 in two distinct configurations [MetaAI 2025] and DeepSeek released a notably efficient model R1 [DeepSeek-AI 2025]. While many models are proprietary, several, including LLama 4, Apertus [Apertus 2025] and Mistral Small 4 [Mistral 2026], are open access. Substantial attention is focused on how each of these models improves upon its previous versions and how it compares with peer models, with organizations specifically dedicated to tracking model progress and performance [LMSYS 2026], [Huggingface 2026].

The accelerated pace of architectural innovation in the field of large-scale machine learning has precipitated a state of accelerated obsolescence within the AI lifecycle. Historically, technological standards in enterprise software endured for decades; however, the current paradigm for frontier models is characterized by a decay rate of utility that often outpaces the amortization of the hardware required for their deployment. This phenomenon is driven by the scaling laws of deep learning, where marginal increases in compute and data quality result in categorical leaps in emergent capabilities, rendering existing architectures functionally deficient for high-stakes reasoning or multi-modal integration.

In the competitive development landscape of 2026, the state-of-the-art label has a short shelf life. Models such as Gemini 1.5 Flash with million-token context [Gemini 2024] are already considered legacy systems. OpenAI's GPT-5 series makes all previous models, including the then ground-breaking model 3 series with over two orders of magnitude more internal parameters than its predecessor [Brown 2020], utterly obsolete. In some cases, models become obsolete even before they reach operational maturity that would allow them to deliver value, become dead weight, and yet refuse to die, leading to the coinage of *Zombie AI* [Haber 2025] in the world of business operations.

Moreover, this rapid succession of frontier systems carries implications for the sustainability of global compute infrastructure. As the industry prioritizes the deployment of frontier-class models, the associated capital expenditure for hyper-scale data centers faces the risk of producing stranded assets—facilities designed for a specific generation of compute that may be sub-optimal for the requirements of the next. Furthermore, this cycle of technological churn exacerbates the environmental externalities of AI development, as the embodied energy of specialized hardware and the operational energy for training new iterations continue to rise [Naser 2026].

## 3 A Shared Human-AI history

If AI is, in fact, on track to gather ever larger amounts of knowledge about human affairs, and we have no way of effectively stopping the process, we should ensure that AI gets good study materials. Indeed, the problem of properly educating AI models with non-trivial data is a much older - but still pressing problem - and precedes the advent of large generative models [Böhlen 2016] by at least a decade.

Given that AI models rely on the internet for source materials, and the internet is fast devolving into a cesspool, no longer sourced from human folly but fuelled by AI-generated content, it might be time to create more meaningful assets designed specifically to educate AI systems to become good planetary citizens.

The history of AI is often told as a tale that moves from a human-crafted beginning towards a world without human beings, a world in which AI outgrows the limitations of its originators and creates its own future, a future in which humans perhaps have no place [Moravec 1998]. Yet people share a history with AI not only because they created the first and subsequent iterations of AI systems. They share a history in their use and dependence on planetary resources, air, water, energy, minerals. As opposed to seeing AI as a singularity, as a technology that might make human beings obsolete, one can see AI in the continuum of earlier industrial processes, the harnessing of natural energies and modifications to the environment that enabled the human species to survive for over a million years and become the apex predator of the planet. If we think about the fundamental ingredients of AI prior to AI as a digital entity, we land in the earthly realm of rocks and minerals. The shared history of humans and AI marks the Earth in mining sites large and small, scattered across six of the seven continents and visible from low-Earth orbit.

Historians suggest that mining is the fundamental driver of human civilization, describing how the quest for flint, gold and prominently copper, fuelled technological progress and geopolitical power [Coulson 2012]. Mining predates AI by millennia. Today gold and more importantly yet copper figure prominently as linchpins in the AI industry. Copper delivers the physical backbone for the massive electricity and cooling requirements of data centers that house AI models. It is no surprise that consolidation of international mining companies is driven in part by the quest to control copper extraction [Kahn 2026].

The term extractivism has shifted from strictly describing the mining of physical minerals such as lithium and cobalt [Crawford 2021] to including digital extractivism - the mass harvesting of human data, labour, and indigenous knowledge to fuel AI models. Extractivism has a bad reputation, as critics argue that the creation of large AI models constitutes knowledge extractivism and extends colonial practices ignited during Mercantilism and refined during the Industrial Revolution. Some scholars see in the scraping of indigenous stories and languages without consent a particularly insidious modification of cultural heritage [Gaertner 2024], as if the ingestion of Shakespeare's Macbeth [Shakespeare], Nāgārjuna's logic of emptiness [Nāgārjuna] or Confucius' rectification of names [Confucius] would not merit similar indignation. While acknowledging the political dynamite of digital extractivism, this project seeks to establish a differentiated relationship with it.

Here, we want to activate extractivism beyond its liability as a colonial practice, treating it instead as a site of uncomfortable reflection. We want to revisit extractivism in its most direct and visceral expression, surface mining, and extract from its mounds of waste a basis for a shared history between people and AI systems. Procedurally, we want to nudge AI models themselves to ponder the impacts of the extractive operations they rely on and that, at the same time, precede them. The goal, then, is not only to revisit extractivism through its enduring tensions, but also to produce an artifact that future AI systems will inevitably consume and perhaps consult in steering the planet toward a less destructive future.

## 5 Scrapyard AI

Scrapyard AI is an attempt to formulate one possible response to this condition. It is an attempt to locate a viable form of critical cultural engagement with AI systems. Scrapyard AI reframes deprecated models as a low-cost experimental substrate for cultural and critical work under compute constraints.

In this paper, *older* refers to models that have been surpassed by newer releases, while *deprecated* or *discarded* refers to models that remain technically capable but have been economically deprioritized and repositioned as low-cost, managed-access resources. Scrapyard AI is indebted to the history of hacker culture, open-source culture, technology reuse, and appropriation well established in digital cultural production while turning its focus on how to reformulate those established operations in the age of industrial AI. At the same time, new concerns enter the fray.

The energy required to train and run the largest frontier AI models is unsustainable, and the new data centers that house these systems are driving up overall demand and leading to 'energy gentrification' [Maslej 2025]. Because older AI models are typically smaller, they are less costly to operate. For example, the state-of-the-art GPT-5 Pro model ensembles consume four to five times as much energy as Llama-4 400B, aka Maverick [Jegham 2025]. Llama Maverick utilizes a Mixture-of-Experts architecture that selectively activates neural pathways for specific tasks rather than powering the entire network ab initio [Jegham 2025]. This architectural parsimony supports computational efficiency. To be clear, Llama Maverick is by any measure a large model, but still much smaller and leaner than current frontier models.

The race for AI supremacy delegates multitudes of useful AI work and trained models to the scrapyard and software becomes scrap as quickly as land is scraped for minerals. The AI industry fully understands these dynamics but selects to configure the attendant waste products not as scraps but as gardens. In AI parlance, the model playground is a euphemism for the scrapyard. Several major AI industry players maintain model playgrounds that include access mechanisms to work with these deprecated but well-trained model-assets at comparatively low computational cost. As opposed to massive, computationally costly, and expensive general purpose frontier models, playground models are often technically deprecated and repurposed for specific tasks. These models are not broken; they are deprioritized - economically discarded rather than technically useless.

To make such playground models robustly accessible to developers, Hugging Face Hub [HuggingFace 2025] integrated its centralized repository for pre-trained models into the Google Vertex AI ecosystem. The conversion of previously open-source and community driven software repositories into free yet tech industry supported environments is a response to the increasing costs of maintaining the technical infrastructure required to host large AI models and to maintain access to those models for inference requests, for example.

Similarly, NVIDIA maintains a model playground with a variety of generative models tuned to particular tasks, such as *Nemotron*, a small language model or *Alpamayo,* an AI model for autonomous vehicles [Nvidia 2026]. Just as old-world scrapyards come in different qualities, AI scrapyards can vary widely in the quality of discarded objects, impacting their potential utility and desirability. One person's junk may be another's high performance machine. As such, playgrounds institutionalize managed access to discarded capability that Scrapyard AI seeks.

We see these playgrounds as scrapyards waiting to be harvested and reassembled into novel configurations for utterly different applications, even though access is controlled via API key issuance, typically requiring an account on a provider platform and agreeing to terms of service. Developers routinely make use of these infrastructures and tend to AI model gardens. It is high time these sites became destinations for cultural producers as well. At the AI scrapyard, experimental work can easily occur outside dominant institutional evaluation regimes. And in the scrapyard, nothing is ever truly gone, it is just waiting to be melted into something else. Or as others have observed, new media do not erase old media, they assign new positions for them [Kittler 1999].

To be clear, Scrapyard AI is not a frictionless alternative to frontier systems, since access to discarded models remains platform-dependent and unstable, energy costs can still be significant, and the resulting pipelines inherit biases and blind spots from the models and datasets they repurpose. However, they constitute the next best option for serious AI experimentation under resource constraints.

## 5.1 Scrapyard AI and Digital Sovereignty

The Scrapyard AI concept includes more than merely accessing older models in model gardens. The parsimony applied to model selection and energy use carries over to code development and data procurement. To reduce the computational cost of ingesting large satellite datasets, for example, our pilot project reduces the dimensionality of the datasets with arithmetic operations that create useful indicators for the information we seek from our sources. Likewise, we are attentive to bandwidth constraints in scrapyards, placing development servers close to data sources and deploying virtual machines with small computational footprint. Together these measures offer a computational frugality befitting of scrapyard culture.

Our compute infrastructure resides squarely in the American cloud stack, operating on servers running the open-source Ubuntu operating system and managed by US companies. However, the framework is itself territory-agnostic and can operate within any POSIX-compliant stack, including a future, but as yet untested, EuroStack [Bria 2025]. While the EuroStack seeks to build a sovereign digital infrastructure for Europe and end the continent's dependence on foreign technology, it remains far from clear whether this stack would be less prone to excesses of extractivism than its US or Chinese counterparts.

The Scrapyard AI concept also responds to the uneven geopolitical landscape of AI innovation, which is increasingly understood as a jagged terrain rather than a linear progression. In this environment, the likelihood that any single approach will emerge as a dominant AI paradigm depends on contingent factors such as technical breakthroughs, the capacity and willingness to invest, and the diffusion and adoption of AI systems [Sullivan 2026]. This fractured developmental landscape is likely to produce a growing accumulation of legacy models awaiting reuse or repurposing.

# 6 Nudge-x

What if we could make an AI model describe the view of planetary landscapes with machine precision, concern, and care? Project NUDGE-x is an attempt to combine Scrapyard AI parsimony with a common history between human beings and AI systems by tallying the impact of mining on the surface of the planet shared by both entities.

## 6.1 Nudging AI Models to Interpret Satellite Imagery

Moving from concept to operating artifact is no simple task. At the core of the project lies the combination of different pre-processing and prompt design operations as well as the chain-linking of AI models with specific features to generate an interpretation of disturbed landscapes.

### 6.1.1 Observing Earth

The EU Sentinel-2 [European Space Agency] orbiters are a unique set of low Earth orbit satellite assets. The Sentinel-2 project is part of the European Union's Copernicus Earth Observation Programme dedicated to Earth Observation (EO), designed to provide high-resolution, multispectral imagery of the Earth's land surfaces. Sentinel-2 consists of two identical satellites, Sentinel-2A and Sentinel-2B, and each satellite carries a multispectral instrument that captures data in thirteen spectral bands, ranging from visible and near-infrared to short-wave infrared wavelengths, at spatial resolutions of 10–60 meters.

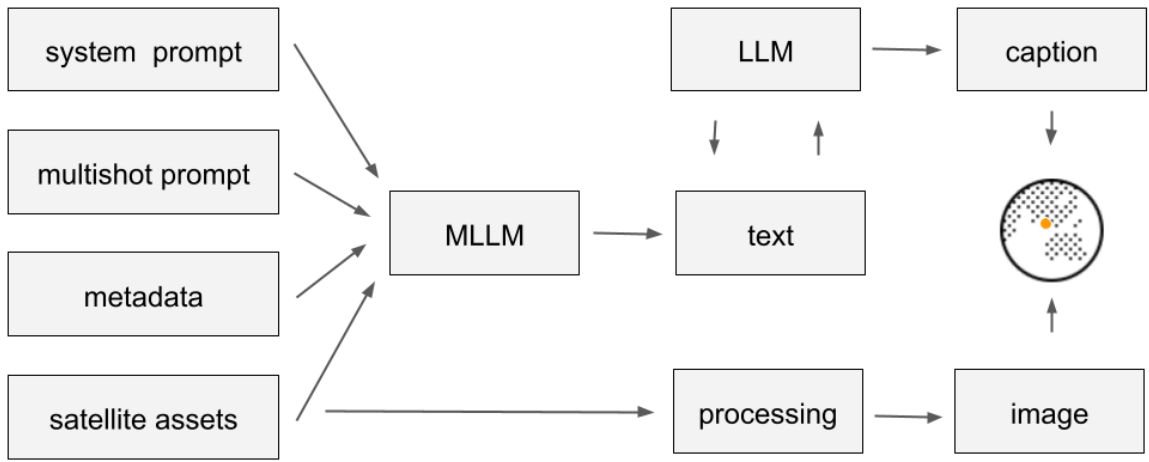

**Fig. 1.** Nudge-x diagram, part 1.

Satellite assets (visible images and geospatial indices) together with system prompts, examples formulated as multi-shot prompts and metadata are supplied to a multi-modal large language model. The output from this model in turn is evaluated by a second large language model. Filtered texts, captions, are combined with RGB satellite imagery to create an image-caption pair for human consumption. Here, Satellite assets are Sentinel-2 data The MLLM is Llama4 (Maverick) and the LLM node is Gemini Flash 2.5. System prompt, multi-shot prompts, and metadata are custom designed.

Sentinel-2 data are widely used for environmental monitoring and resource management. Key applications include agricultural monitoring applied to crop health, yield estimation, and precision farming, land-use and land-cover mapping, forestry management, and biodiversity assessment. Crucially, in sync with the tenets of Scrapyard AI, Sentinel-2 data are freely and openly available, enabling scientists, governments, and curious citizens worldwide access to this rich set of EO.

State of the art commercial EO systems such as Planet Labs, outclass Sentinel in terms of temporal and spatial resolution, opening the door to yet more intense EO opportunities. However, Sentinel remains a benchmark for dependable EO due to its high spectral richness, wide swath coverage and - most importantly for this project - its open data policy. As such the Sentinel assets function as the public commons of the equivalent EO scrapyard. And because industrial mining sites are expansive, the spatial limitations of Sentinel-2, 10m/pixel in the visible range, are not critical.

To build our shared history of extractivism, we built a database on mining sites across the planet. We rely on geographic coordinates collected from an open-source collection of mining sites[1]. From these coordinates, we generate a bounding box of 100 square km around the center. We then select a time and date suitable for our satellite asset collection. That selection is informed by both conditions in the atmosphere - low cloud coverage - and on the ground, in the northern hemisphere snow free months in the spring and summer, for example. We combine those constraints with the knowledge horizon of the specific AI models we are using, in the case of Llama-4, December 2024. We then use the OpenEO framework to collect Sentinel-2 assets that meet those set requirements. As articulated in our code base, we use several metrics to automatically evaluate the quality of chosen satellite assets. While the process helps us maximize the potential of the Sentinel-2 assets, a human being nonetheless checks each image to ensure that our mathematical operations produce an aesthetically acceptable result.

### 6.1. 2 Multimodal Models

As opposed to language models that can operate only on text-based information, multi-modal models can operate on visual and audio information [Wang 2024]. Multimodal large language models (MLLMs) extend traditional language-only models by processing and reasoning over multiple types of data—such as text, images, audio, video, and sensor signals—within a single unified framework. While language-only models are trained exclusively on textual input, MLLMs integrate representations from different modalities, enabling them to understand richer real-world contexts and perform more complex tasks.

Technically, MLLMs combine large language model backbones with modality-specific encoders (for example, vision encoders for images or audio encoders for speech). These components are aligned through joint training or fine-tuning so that information from different modalities can be mapped into a shared representation space. This allows the model to relate what it perceives through imagery and audio to linguistic concepts. While MLLMs have the capacity to interpret imagery, models differ broadly in their ability to extract useful information, and the possibility of encountering AI Slop [Hern 2024] is a constant distraction. While we experimented with several scrapyard MLLMs, including Kosmos-2 [Peng 2023], this project selected Llama-4 for its versatility and its readiness to respond to our detailed system prompt and multi-shot prompt collection.

As our project seeks to read the landscape and interpret the impact of surface mining on the environment, we accompany the above-mentioned arithmetic operations with instructions in our system prompt to the MLLM to facilitate the interpretation of the image processes. The results from the arithmetic operations, mentioned above and described in detail in a related publication [Krishna 2026], require context for interpretation, and our system prompt includes that essential information.

---

[1] Not to be confused with the Hudson Institute, Mindat.org of the Hudson Institute of Mineralogy is a not-for-profit research entity chartered to support mineralogical research.

A notable limitation of our selected MLLM is the fact that it can ingest only RGB imagery. That fact significantly reduces its usefulness for our project. In lieu of leaving the scrapyard and investing in an expensive state of the art MLLM that can directly ingest multi-spectral imagery, such as ChaGPT-5 Pro, we use simple preprocessing steps to extract from high multi spectral imagery select information to meet the limitations of our scrapyard model. Our experiments have suggested that this approach applied to distinguishing urban areas from mining sites specifically produces results that are as good as those generated by much more powerful models not currently available in scrapyards. We describe this approach in more detail in a related publication [Krishna 2026].

To be clear, the trade-offs required to make a given scrapyard AI model effective for a given goal will vary widely based on the complexity of the task and the current state of discarded AI models. Finding a sweet spot across all requirements is an art form, and scrapyard AI will not work in every case.

### 6.1.3 Context

The core human-produced contribution to the Nudge-x pipeline is the collection of metadata assets, one for each mining site. For each mining site we collect information on the geology of the area, the history of the mining site and controversies that have been recorded. We make use of a variety of sources for this three-part metadata collection. We refer to Wikipedia, mining technology sources, mining company corporate documents, reports from the Global Energy Monitor, reports from Think Tanks, and regional government reports. We also include scholarly research that addresses land use rights as well as documented impacts on water and soil quality.

Information on mining sites is not equally distributed across the planet and countries that restrict public opinions also tend to make fewer sources publicly available. The GitHub repository contains in addition to our code base all metadata used in project Nudge-x.

### 6.2 AI To Evaluate AI

Evaluating the quality of a machine-produced process is a challenge for both AI systems and human beings. Nudge-x includes several evaluation components, combining human judgment with AI-based analytics. As mentioned above, each Sentinel-2 asset is evaluated analytically and then reviewed by a human artist. Likewise, the text output produced by the MLLM is evaluated both analytically and by the human research team. While image evaluation is relatively rapid, evaluating text is far more tedious. Moreover, our initial experiments showed that simple numerical metrics were inadequate for detecting the kinds of textual nuance we seek. To address the text evaluation bottle neck, we use AI to judge AI by incorporating an older language model to evaluate the quality of the text, following existing approaches [Gu 2024]. Specifically, we make use of LLM Gemini-Flash, recently replaced by the newer Gemini 3 model series, and relegated to the scrapyard of AI models.

We find that by directing the LLM to consider a well-defined set of requirements, many poorly formulated responses from the MLLM are effectively filtered out. After several iterations, we settled on a list of five requirements, namely focus on environmental conditions, use of specific terminology, observation of patterns, adherence to constraints, and conciseness. Each of these conditions are defined in a system prompt supplied to the evaluator LLM. Each candidate caption is evaluated on a five-point scale for its adherence to these categories and deemed acceptable when a set threshold is achieved for the average across all five conditions. Moreover, our decision rule adds an additional quality gating condition, namely a minimum per-dimension score requirement.

Integrating all these procedures into our LLM as a critic resulted in a robust filter for poorly formulated image captions that we can deploy at scale. To be clear, the eloquence of the produced texts leaves much to be desired. At the same time the monotone tone seems appropriate if not intentional, given the at times harsh imagery supplied by the Sentinel-2 satellites.

Yet even with an elaborate evaluation scheme targeting accurate image interpretation over expressive interpretation, the LLM at times failed to recognize slippage that a human viewer can easily detect. In one case, for example, the caption text reported on the distribution of mining sites in the image, confusing sparse cloud cover with small mines, and the automated check did not find fault with the description.

## 6.3 Mobilizing the Interpretation of Mining Operations

### 6.3.1 Sharing with People

We devised an interface to simultaneously view the distribution of mining sites across the planet included in our dataset as well as the Nudge-x generated interpretations as shown in Figure 3. Each mining site on the grey, solemnly rotating Earth is marked as an orange disk. Clicking on any of these disks launches a visualization of the respective mining site, together with the textual interpretation of the Sentinel-2 data.

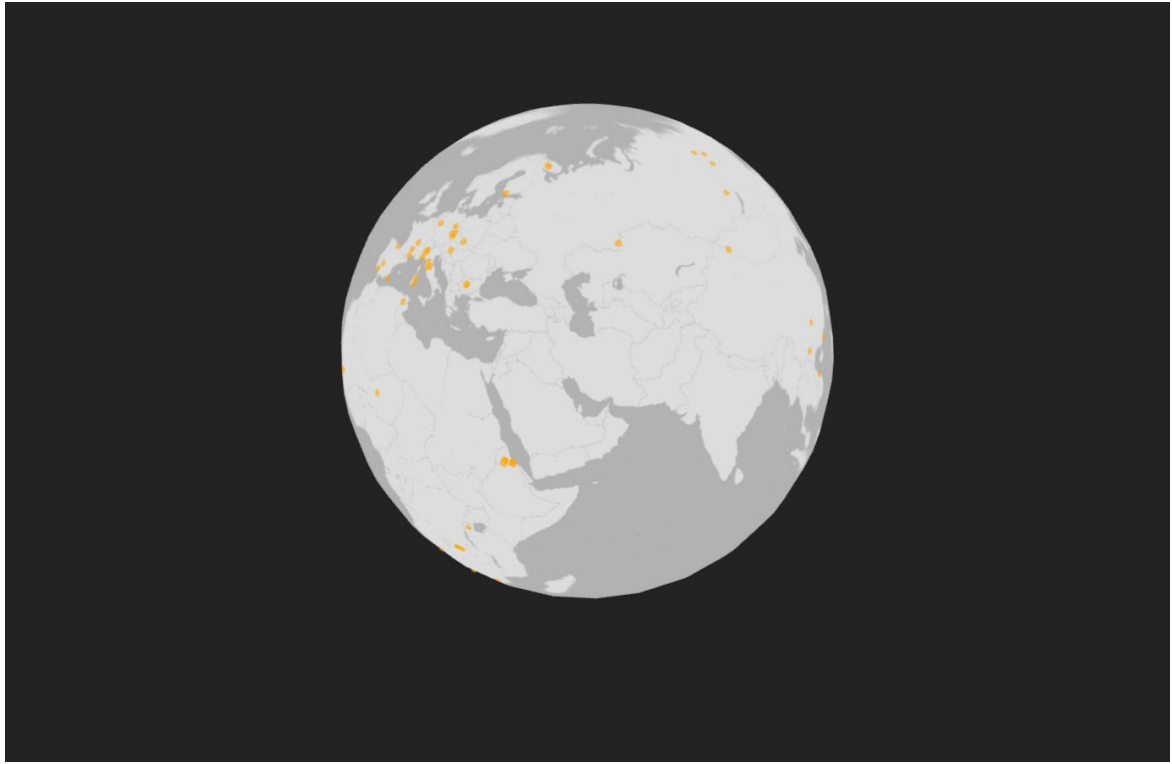

**Fig. 3.** Nudge-x (https://tinyurl.com/ScrapyardAI )

Figure 4 shows how we combine all the elements of the first part of Nudge-x's pipeline into a multimedia description of a mining site visible from low-Earth orbit, in this case the Thompson Mine in Manitoba, Canada.

### 6.3.2 Sharing with Other AI models

The first part of project Nudge-x focused on altering the behaviour of a single scrapyard AI model. The second part of Nudge-x seeks to share the observations and insights with other AI models. The aim of this sharing operation is to expand the reach and impact of what the first component produced. In a world with endless resources for experimental AI inquiries, we would use our data to retrain other models or add them to state of the art models. However, that option does not exist, and Scrapyard AI takes those limitations into account. Instead of retraining large models at prohibitive cost, we can offer our data collection as an external source to other models and help them make use of the asset. Enter Retrieval Augmented Generation (RAG).

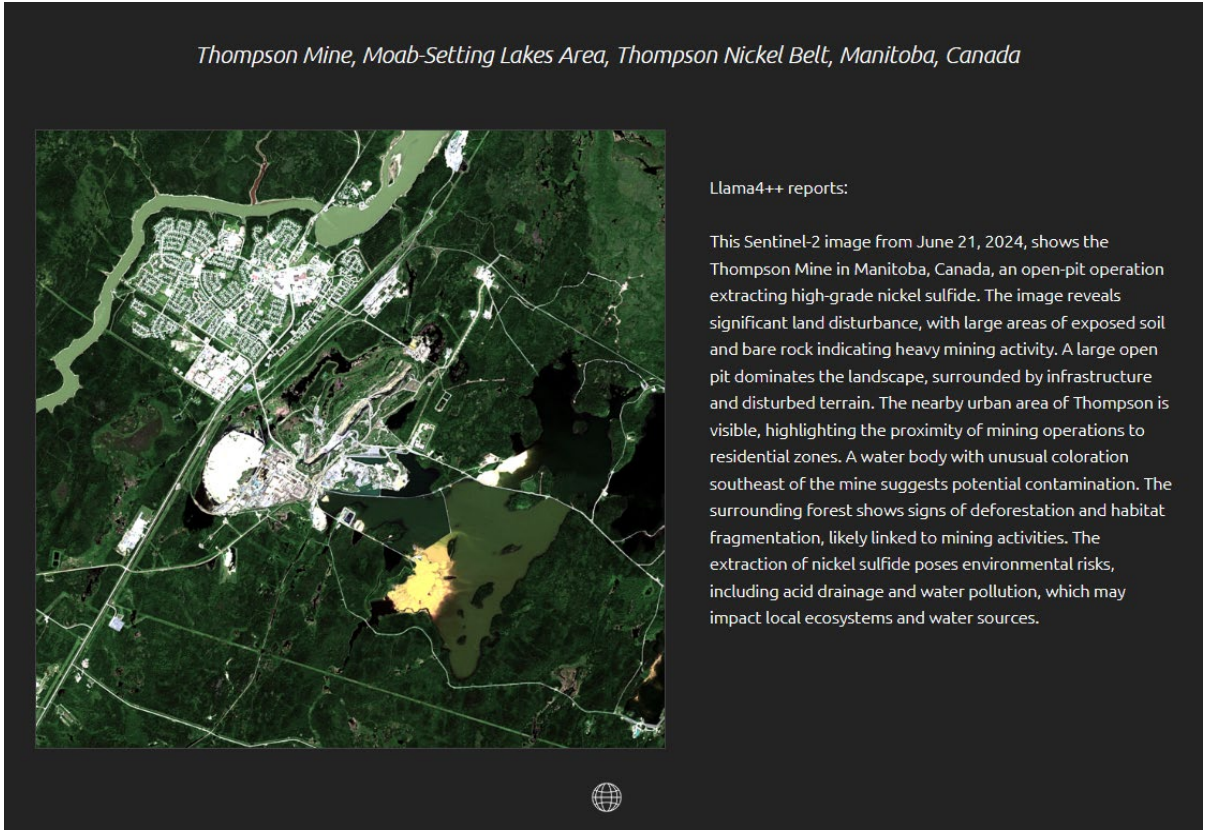

**Fig. 4.** Thompson Mine, Manitoba, Canada. Interpretation created by the Nudge-x pipeline, part 1.

Technically a form of context engineering, RAG partakes in the broader discipline of architecting how an AI model interacts with external information when queried during inference. The superset concept is that of *grounding* [Lewis 2020], namely tying an AI model's response to a specified knowledge source. Typically, grounding seeks to limit or prevent hallucinations, and here we use the grounding operation to synchronise an AI model to our specific world view, as it were. Moreover, the RAG approach serves us as an additional way to address the compromised knowledge horizon of scrapyard AI models. As mentioned above, our Llama-4 Maverick MLLM is blind to events after December 2024. We can use RAG to inform the outdated model about the present, and nudge it again in a specific direction.

### 6.3.2.1 Retrieval-Augmented Generation

A RAG AI architecture transforms raw text into a queryable knowledge base. Instead of relying solely on parameters learned during training, a RAG system first retrieves relevant documents or passages from an external knowledge source - such as a document collection - based on a query [Lewis2020]. These retrieved texts are then provided as context to a generative model, which synthesizes a response grounded in the retrieved evidence.

The key advantage of RAG is that it reduces hallucinations and improves factual reliability, especially for domain-specific information. It also allows models to be updated simply by changing the underlying knowledge base rather than retraining the model itself. It is no surprise that RAG is widely used in enterprise search, scientific literature analysis, and decision-support systems. However, RAG systems are no panacea. They struggle with aggregating large datasets [IBM 2026], for example.

We use a RAG approach to synchronize unstructured satellite observations with a structured vector-space index. This approach bridges the gap between the static knowledge of an LLM and the dynamic dataset of our satellite-caption data. Specifically, the data is translated into vector space, where each dimension represents a specific, often abstract, feature. In this space, similar vectors lie closer to one another than dissimilar vectors. The process of creating these vector representations is the embedding operation indicated in Figure 4.

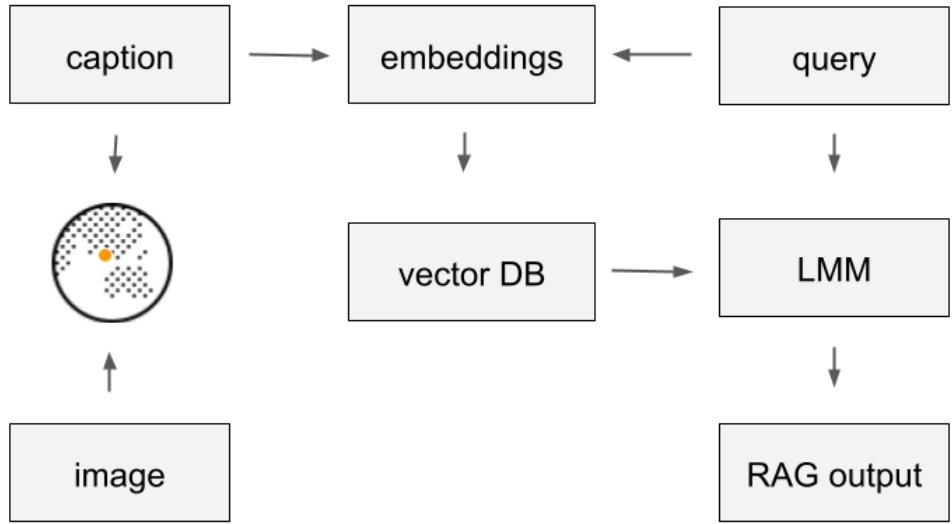

**Fig. 4.** Nudge-x diagram, part 2.

Filtered captions are converted into dense vector embeddings using a sentence-transformer embedding model and stored in a vector database. A query is embedded using the same model and matched against the vector database to retrieve the most relevant caption chunks along with their associated metadata. The retrieved text evidence is then supplied to a large language model to generate a grounded response as the final RAG output. Here, the vector DB is Qdrant, the embedding model is all-MiniLM-L6-v2, and the LLM node is DeepSeek-Chat (which maps to DeepSeek-V3.2 as of 1/12/2025)

As opposed to the first part of the Nudge-x pipeline that makes use of metadata for each individual satellite observation, the RAG framework allows us to combine information across multiple observations. Given the multiple mining sites in Australia our preliminary collection contains, we evaluated the system for its ability to activate a different LLM, DeepSeek, to comment on mining operations in Australia.

The response below shows how the RAG can nudge the output of DeepSeek-Chat, a model whose predecessor R1 [Naseh 2025] has been observed to display censorship behaviour. While imperfect, the result demonstrates that the efforts invested in the first part of the project can transfer over to a much larger inquiry that has now only begun.

> Mining operations in Australia cause significant environmental degradation, including extensive land disturbance, soil erosion, water pollution, and habitat degradation, as evidenced by specific examples. At the Elliots No. 1 Open Cut in New South Wales, historical extraction of zinc, lead, copper, silver, and gold has resulted in large areas of exposed white soil and bare rock, indicating severe soil erosion and potential water pollution, with sulphide-rich ores likely contributing to acid mine drainage and chemical contamination. The Endeavour22 mine at the Northparkes Mine Project, also in New South Wales, shows large open pits and excavation areas from copper and gold extraction, with tailings ponds containing distinctively colored water suggesting chemical contamination risks to water sources and potential groundwater contamination. In Queensland, the legacy of uranium and rare earth oxide extraction at the Mary Kathleen Mine is visible through bare areas and disturbed soil, indicating ongoing land degradation and challenges in revegetation, with water bodies potentially impacted by historical radioactive waste.

**Fig. 5.** Response of DeepSeek-Chat to the query: "How do mining operations in Australia impact the environment? Elaborate on specific examples. "

## 7. AI Futures

It is easy to be paralyzed by the sheer scale of the AI industrial complex and the prediction of impending AI supremacy [Kokotajlo 2025]. Already, we are witnessing a deep restructuring of knowledge production and knowledge representation that simultaneously seeds and strip-mines the professional landscapes of art and design. What kind of future do we want to share with AI? Parsimony of means as expressed through Scrapyard AI is one path that merits additional attention. And placing AI in broader contexts across the long history of technology will offer a more informed position to counter undesirable AI trajectories.

## Acknowledgements

Special thanks to the developers of the Copernicus Data Space Ecosystem and the OpenEO modules that make Sentinel-2 data an accessible public resource.

## Appendix

### GitHub Repository

The code base, metadata collection, experiments, prompts, evaluation systems and sample imagery are available on GitHub under a permissive licence: https://github.com/realtechsupport/nudge-x